\documentclass[10pt,openany]{book}
\usepackage{natbib}
\usepackage{graphicx}
\usepackage{color}
\usepackage{hyperref}
\usepackage{enumitem}
\usepackage{enumerate}
\usepackage{nopageno}

\definecolor{mygreen}{rgb}{0.19,0.55,0.11}
\definecolor{dkgreen}{rgb}{0,0.6,0}

\makeindex
\begin{document} 

\pagenumbering{arabic}

\cleardoublepage
\thispagestyle{empty}

\begin{center}
{\Huge Physics of Binary Star~Evolution}\\
\vspace{0.5cm}
{\large --- from Stars to X-ray Binaries and Gravitational Wave Sources}

\vspace{3.0cm}
{\Large Thomas M. Tauris}

\vspace{1.0cm}
{\Large Edward P.J. van den Heuvel}

\vspace{6.0cm}
{\large Princeton University Press}


\vspace{0.5cm}
{\large Authors' \LaTeX version of textbook in press --- May, 2023}

\vspace{2.0cm}
{\normalsize {\color{blue}Direct link to textbook at publisher:\\} \url{https://press.princeton.edu/books/hardcover/9780691179070/physics-of-binary-star-evolution}}

\vspace{1.3cm}
{\normalsize {\color{blue}For any citations to this textbook, please use the ADS entry:\\} \url{https://ui.adsabs.harvard.edu/abs/2023pbse.book.....T/abstract}}

\end{center}


\setcounter{chapter}{0} 
\thispagestyle{plain}
\chapter*{Contents}
\label{chap:arXiv}

{\bf Preface} \dotfill xi\\
\\
{\bf 1 Introduction: The Role of Binary Star Evolution in Astrophysics}  \dotfill 1\\
\\
{\bf 2 Historical Notes on Binary Star Discoveries}  \dotfill 10

2.1 Visual Binaries and the Universal Validity of the Laws of Physics \dotfill 10

2.2 Astrometric Binaries \dotfill 11

2.3 Spectroscopic Binaries \dotfill 14

2.4 Eclipsing Binaries \dotfill 15

2.5 The Discovery of the Binary Nature of Novae and Other Cataclysmic \\ \indent \indent Variables \dotfill 17

2.6 The Discovery of the Binary Nature of the Brightest X-ray Sources \\ \indent \indent in the Sky \dotfill 20

2.7 Centaurus X-3: Discovery of the First Neutron Star X-ray Binary \dotfill 21

2.8 Cygnus X-1: Discovery of the First Black Hole X-ray Binary \dotfill 22

2.9 The Discovery of the Existence of Double NSs and Double BHs \dotfill 25

2.10 The Discovery of Millisecond Radio Pulsars: Remnants of LMXBs \dotfill 26

2.11 Type Ia, Ib, and Ic SNe: Results of the Evolution of Binary Systems \dotfill 27

2.12 Binary Nature of Blue Stragglers, Barium Stars, and Peculiar \\ \indent \indent Post-AGB Stars \dotfill 29

Exercises \dotfill 31\\
\\
{\bf 3 Orbits and Masses of Spectroscopic Binaries} \dotfill 33

3.1 Some Basics about Binary Orbits \dotfill 33

3.2 Orbit Determination \dotfill 36

3.3 Determination of Stellar Masses \dotfill 41

3.4 Masses of Unevolved Main-sequence Stars \dotfill 42

3.5 The Most Massive Stars \dotfill 44

3.6 Falsification of Radial Velocity Curves \dotfill 46

3.7 The Incidence of Interacting Binaries and Their Orbital \\ \indent \indent Distributions and Masses \dotfill 51

Exercises \dotfill 57\\
\vspace{2cm}
\\
{\bf 4 Mass Transfer and Mass Loss in Binary Systems} \dotfill 59

4.1 Roche Equipotentials \dotfill 59
 
4.2 Limitations in the Concept of Roche Equipotentials \dotfill 63

4.3 Orbital Changes due to Mass Transfer and Mass Loss \\ \indent \indent in Binary Systems \dotfill 65

4.4 Observational Examples \dotfill 83

4.5 Basic Physics of Mass Transfer via $L_1$ \dotfill 88

4.6 Accretion Disks \dotfill 98

4.7 Tidal Evolution in Binary Systems \dotfill 109

4.8 Common Envelopes \dotfill 115

4.9 Eddington Accretion Limit \dotfill 131

Exercises \dotfill 134\\
\\
{\bf 5 Observed Binaries with Non-degenerate or \\ \indent White Dwarf
Components} \dotfill 139

5.1 Introduction \dotfill 139

5.2 Unevolved Systems  \dotfill 142

5.3 Evolved Systems with Non-degenerative Components \dotfill 143

5.4 Systems with One or Two White Dwarfs \dotfill 152
  
Exercises \dotfill 167\\
\\
{\bf 6 Observed Binaries with Accreting Neutron Stars and \\ \indent Black Holes: X-ray Binaries} \dotfill 168

6.1 Discovery of NS and BH Character of Bright Galactic X-ray Sources \dotfill 168

6.2 Two Types of Persistent Strong X-ray Sources: HMXBs and LMXBs \dotfill 176

6.3 HMXBs and LMXBs vs. IMXBs \dotfill 180

6.4 Determinations of NS Masses in X-ray Binaries \dotfill 188

6.5 BH X-ray Binaries \dotfill 191

6.6 Binaries and Triples with Non-interacting BHs \dotfill 206

Exercises \dotfill 209\\
\\
{\bf 7 Observed Properties of X-ray Binaries in More Detail} \dotfill 213

7.1 High-mass X-ray Binaries in More Detail \dotfill 213 

7.2 Stellar Wind Accretion in More Detail \dotfill 227

7.3 Spin Evolution of Neutron Stars \dotfill 232

7.4 The Corbet Diagram for Pulsating HMXBs \dotfill 243

7.5 Orbital Changes due to Torques by Stellar Wind Accretion, \\ \indent \indent Mass Loss, and Tides \dotfill 245

7.6 Measuring BH Spins in X-ray Binaries \dotfill 245

7.7 Ultra-luminous X-ray Binaries \dotfill 252

7.8 Low-mass X-ray Binaries in More Detail \dotfill 258

Exercises \dotfill 270\\
\vspace{1cm}
\\
{\bf 8 Evolution of Single Stars} \dotfill 271

8.1 Overview of the Evolution of Single Stars \dotfill 271

8.2 Final Evolution and Core Collapse of Stars More Massive \\ \indent \indent than $8\;M_\odot$ \dotfill 299

8.3 Evolution of Helium Stars \dotfill 315

Exercises \dotfill 325\\
\\
{\bf 9 Stellar Evolution in Binaries} \dotfill 326

9.1 Historical Introduction: Importance of Mass Transfer \dotfill 326

9.2 Evolution of the Stellar Radius and Cases of Mass Transfer \dotfill 327

9.3 RLO: Reasons for Large-scale Mass Transfer and Conditions for \\ \indent \indent Stability of the Transfer \dotfill 334

9.4 Results of Calculations of Binary Evolution with \\ \indent \indent Conservative Mass Transfer \dotfill 340

9.5 Examples of Non-conservative Mass Transfer \dotfill 353

9.6 Comparison of Case B Results with Some Observed Types \\ \indent \indent of Systems \dotfill 360

9.7 Differences in Final Remnants of Mass-transfer Binaries and \\ \indent \indent Single Stars \dotfill 366

9.8 Slowly Rotating Magnetic Main-sequence Stars: \\ \indent \indent The Products of Mergers? \dotfill 371

Exercises \dotfill 374\\
\\
{\bf 10 Formation and Evolution of High-mass X-ray Binaries} \dotfill 376

10.1 Introduction: HMXBs are Normal Products of \\ \indent \indent Massive Binary
Star Evolution \dotfill 376 

10.2 Formation of Supergiant HMXBs \dotfill 376

10.3 Formation of B-emission (Be)/X-ray Binaries \dotfill 379

10.4 WR Binaries, HMXBs, and Runaway Stars \dotfill 386

10.5 Stability of Mass Transfer in HMXBs \dotfill 393

10.6 The X-ray Lifetime and Formation Rate of the \\ \indent \indent $\;$ Blue Supergiant HMXBs \dotfill 395

10.7 Highly Non-conservative Evolution and Formation of \\ \indent \indent $\;$ Very Close Relativistic Binaries \dotfill 403

10.8 Formation Models of HMXBs Different from Conservative \\ \indent \indent $\;$ Case B Evolution \dotfill 408

10.9 The Lower Mass Limit of Binary Stars for Terminating as a BH \dotfill 411

10.10 Final Evolution of BH-HMXBs: Two Formation Channels for \\ \indent \indent $\;$ Double BHs \dotfill 414

10.11 Final Evolution of Wide-orbit BH-HMXBs via CE Evolution \dotfill 415

10.12 Final Evolution of Relatively Close-orbit BH-HMXBs \\ \indent \indent $\;$ via Stable RLO \dotfill 419

10.13 Reﬁnement of the DNS Formation Model: Case BB RLO in \\ \indent \indent $\;$ Post-HMXB Systems \dotfill 423

\enlargethispage{1\baselineskip}
Exercises \dotfill 431\\
\vspace{1cm}
\\
{\bf 11 Formation and Evolution of Low-mass X-ray Binaries} \dotfill 433

11.1 Origin of LMXBs with Neutron Stars \dotfill 433

11.2 Origin of LMXBs with Black Holes \dotfill 449

11.3 Mechanisms Driving Mass Transfer in Close-orbit LMXBs
and CVs \dotfill 450

11.4 Formation and Evolution of UCXBs \dotfill 464

11.5 Mechanisms Driving Mass Transfer in Wide-orbit LMXBs \\ \indent \indent $\;$ and Symbiotic Binaries \dotfill 470

11.6 Stability of Mass Transfer in Intermediate-Mass and \\ \indent \indent $\;$ High-Mass X-ray Binaries \dotfill 475

Exercises \dotfill 477\\
\\
{\bf 12 Dynamical Formation of Compact Star Binaries in \\ \indent \, Dense
Star Clusters} \dotfill 480

12.1 Introduction \dotfill 480

12.2 Observed Compact Object Binaries in Globular Clusters: \\ \indent \indent $\;$ 
X-ray Binaries and Radio Pulsars \dotfill 482

12.3 Possible Formation Processes of NS Binaries in Globular Clusters \dotfill 483

12.4 Dynamical Formation of Double BHs \dotfill 489

12.5 Compact Objects in Globular Clusters Constrain Birth Kicks \dotfill 492\\
\\
{\bf 13 Supernovae in Binaries} \dotfill 495

13.1 Introduction \dotfill 495

13.2 Supernovae of Type Ia \dotfill 498

13.3 Stripped-Envelope Core-Collapse SNe \dotfill 513
 
13.4 Electron-capture SNe in Single and Binary Stars \dotfill 518

13.5 Ultra-Stripped Supernovae \dotfill 523

13.6 Comparison between Theory and Observations of SNe Ib and Ic \dotfill 529

13.7 Supernova Kicks \dotfill 531

13.8 Kinematic Impacts on Post-SN Binaries \dotfill 541

Exercises \dotfill 556\\
\\
{\bf 14 Binary and Millisecond Pulsars} \dotfill 560

14.1 Introduction to Radio Pulsars \dotfill 561

14.2 To Be Recycled or Not to Be Recycled \dotfill 571

14.3 MSPs with He WD or Sub-stellar Dwarf Companions \\ \indent \indent $\;$ –- Evolution from LMXBs \dotfill 578

14.4 MSPs with CO WD Companions–Evolution from IMXBs \dotfill 591

14.5 Formation of MSPs via Accretion-induced Collapse \dotfill 595

14.6 Recycling of Pulsars \dotfill 597

14.7 Masses of Binary Neutron Stars \dotfill 618

14.8 Pulsar Kicks \dotfill 635

14.9 Formation of Double Neutron Star Systems \dotfill 637

Exercises \dotfill 648\\
\vspace{2cm}
\\
{\bf 15 Gravitational Waves from Binary Compact Objects} \dotfill 652

15.1 The Evidence of GWs prior to LIGO \dotfill 655

15.2 GW Luminosity and Merger Timescale \dotfill 658

15.3 Observations of GW Signals from Binaries \dotfill 661

15.4 Galactic Merger Rates of Neutron Star/Black Hole Binaries \dotfill 664

15.5 Formation of Double Black Hole Binaries \dotfill 667

15.6 Properties of GW Sources Detected so Far \dotfill 678

15.7 Empirical Merger Rates \dotfill 694

15.8 BH Spins–Expectations and Observations \dotfill 696

15.9 Anticipated Other Sources to be Detected in the GW Era \dotfill 706

15.10 GW Follow-up Multimessenger Astronomy \dotfill 712

15.11 Cosmological Implications \dotfill 718

15.12 LISA Sources \dotfill 718

15.13 LISA Sensitivity Curve and Source Strain \dotfill 730

Exercises \dotfill 736\\
\\
{\bf 16 Binary Population Synthesis and Statistics} \dotfill 739

16.1 Introduction \dotfill 739

16.2 Methodology of Population Synthesis \dotfill 741

16.3 Empirical vs. Binary Population Synthesis-Based Estimates of \\ \indent \indent $\;$ Double Compact Object Merger Rates \dotfill 747

16.4 Some History of Early Binary Population Synthesis: \\ \indent \indent $\;$ Evolution of Open Star Clusters with Binaries \dotfill 753\\
\\
{\bf Acknowledgments} \dotfill 761\\
\\
{\bf Answers to Exercises} \dotfill 765\\
\\
{\bf List of Acronyms} \dotfill 767\\
\\
{\bf References} \dotfill 771\\
\\
{\bf Index} \dotfill 843\\

\cleardoublepage
\chapter*{Preface}
\label{chap:preface}
\addcontentsline{toc}{chapter}{Preface}
\thispagestyle{empty}
The majority of all stars are members of a binary system. 
The evolution of such binary stars and their subsequent production of pairs of compact objects in tight orbits, such as double neutron stars and double black holes, play a central role in modern astrophysics, Binary evolution leads to the formation of different types of violent cosmic events such as novae, supernova explosions, gamma-ray bursts, mass transfer and accretion processes in X-ray binaries, and the formation of exotic radio millisecond pulsars. In some cases, the binary systems terminate as spectacular collisions between neutron stars and/or black holes. These collisions lead to powerful emission of gravitational waves, as detected by LIGO since 2015. The coming decade is expected to reveal a large number of discoveries of binary compact systems, as well as their progenitors and merger remnants, from major instruments such as the radio Square-Kilometre Array; the gravitational wave observatories LIGO--Virgo--KAGRA--IndIGO and LISA; the astrometric space observatory Gaia; the James~Webb Space Telescope; and the X-ray space observatories eXTP, STROBE-X, and Athena. In this light, it is important to have a modern textbook on the physics of binary stars evolution, from ordinary stars to X-ray binaries and gravitational wave sources. 
The scope of this book is that the reader (student or educated expert) will learn the physics of binary interactions, from stellar birth to compact objects, and relate this knowledge to the latest observations. The reader will learn about stellar structure and evolution, and detailed binary interactions covering a broad range of phenomena, including mass transfer and orbital evolution, formation and accretion onto compact objects (white dwarfs, neutron stars and black holes), and their observational properties. Exercises are provided throughout the book.

\vspace{0.5cm}
\noindent It has been a privilege for us to write this book, and we hope you enjoy it.

\vspace{1.5cm}
\begin{tabbing}
Thomas Tauris $\;\;\;\;\;\;\;\;\;\;\;\;\;\;\;\;\;\;\;\;\;\;$\&$\;\;\;\;$ \= Ed van den Heuvel \hspace{0.5cm}--- \hspace{0.1cm}June~2022\\
(Aalborg / Aarhus / Bonn) \> (Amsterdam)
\end{tabbing}

\newpage
\begin{figure*}[!t]
\vspace{0.5cm}
  \begin{center}
  \includegraphics[width=0.95\textwidth, angle=0]
{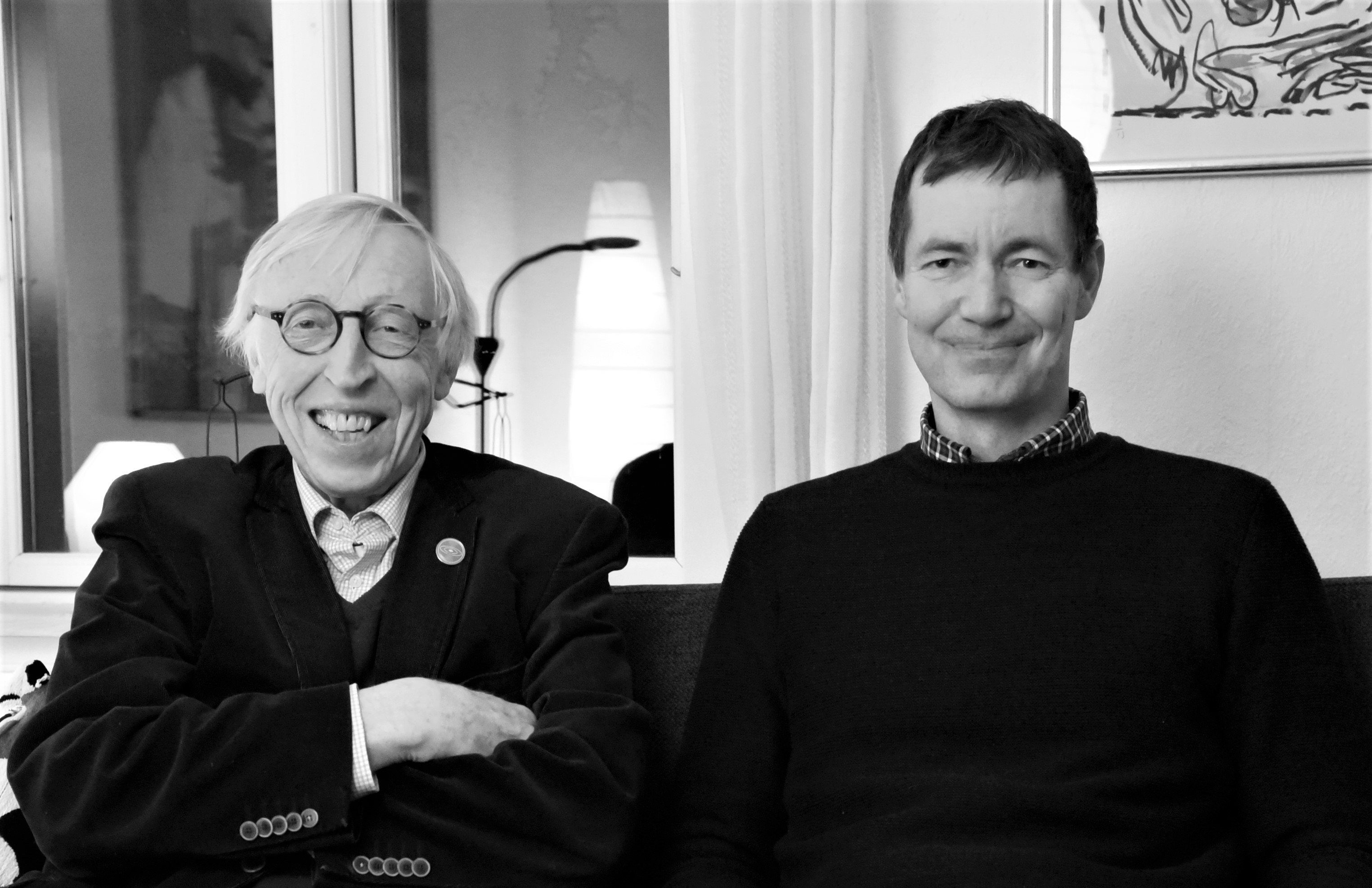}
\caption{The authors (Ed van den Heuvel and Thomas Tauris) enjoying a relaxed evening in Thomas' house in Djursland, Denmark, after a long day's work on the book in February 2020. [Photo by Birgitte Tauris.]}
  \label{fig:authors}
  \end{center}
\end{figure*}

\cleardoublepage
\thispagestyle{plain}
\chapter[Introduction]{Introduction:\\ The Role of Binary Star Evolution in Astrophysics}
\label{chap:intro}

Many key astrophysical objects and phenomena are related to the evolution of binary stars. This holds, for example, for the formation of the brightest \linebreak X-ray sources in the sky and the formation of double neutron stars (NSs) and black holes (BHs), the mergers of which produce the strongest bursts of energy anywhere in the observable Universe, measurable on Earth with gravitational wave (GW) detectors\footnote{The first ever detected BH merger event GW150914 \citep{aaa+16a} released an energy of $\sim\!3\;M_\odot\,c^2$ within a fraction of a second, thereby outshining all stars in the Universe for a brief moment.}. Binary interactions also play a role for the origin of most supernovae (SNe), all nova explosions and short gamma-ray bursts, and the formation of millisecond radio pulsars and a large variety of stars with peculiar chemical abundances, such as barium stars and carbon-enriched metal poor stars.

It has been long realized, from the fact that most stars are members of binary systems \citep{abtl76,bbv03}, that binary evolution must play a key role in stellar evolution \citep[e.g.][]{vdh94a}. 
This early and important awareness has even been strengthened further by the findings of e.g. \citet{cnh+11,sdd+12}, that practically all massive stars are found in binaries with orbits such that at some stage in their evolution, the far majority of these stars will interact with each other. This implies that binary interactions dominate the evolution of massive stars. 

The first ideas about the evolution of binary systems originated in the 1950s. They were largely inspired by the surprising characteristics of Algol-type eclipsing binary systems (see also Section~2.4). Here, {\em Algol-type} means physically similar to the Algol-system, which consists of an unevolved B8V main-sequence star with a mass of $3.2\;M_\odot$ together with an evolved but less massive subgiant companion star of spectral type G8\,III, of mass $0.7\;M_\odot$. 
This situation, with the more evolved star having the smaller mass of the two, is just opposite to what one would expect on the grounds of stellar evolution, as stars of larger mass are expected to have shorter lives than stars of smaller mass do. In a binary --- where both stars were born at the same time --- one would therefore expect the star of larger mass to be in a more advanced stage of evolution at any time than that of its companion of smaller mass.

This is what is called the {\em Algol paradox}. \citet{cra55} was the first to realize that this paradoxical situation can be explained if one assumes that large-scale mass transfer can take place during the evolution of a binary system: Crawford hypothesized that the subgiant components in Algol-type binaries were originally the more massive components of these systems. As the more massive star evolved faster than its less massive companion did, it was the first one to exhaust the hydrogen fuel in its core and evolve into a giant star with a much expanded envelope. The presence of the close companion, however, prevented such an evolution: when the outer layers of the expanding (sub)giant came under the gravitational influence of the companion, they were captured by this smaller star (the {\em accretor}), causing the accretor to increase in mass at the expense of the (sub)giant (the {\em donor}). The (sub)giant transferred so much of its mass that it was finally able to restabilize its internal structure. At that moment, it had become the less massive of the two stars, and the originally less massive star became the most massive one of the pair. 

The first attempt to carry out a real calculation of this type of evolution with mass transfer was by \citet{mor60}. He demonstrated the correctness of Crawford's conjecture, that mass transfer, once it begins, continues until the (sub)giant has become the less massive star of the system. In his calculations, however, Morton still assumed that the orbital period of the system does not change during the mass transfer. 
This is not correct because, if one assumes that the total mass of the system is conserved, one also expects the total angular momentum of the system to be conserved. The total angular momentum is, in good approximation, equal to the orbital angular momentum (as the rotational angular momentum of the two stars is usually much smaller than the orbital one). Conservation of the orbital angular momentum implies that, during the mass transfer the orbital period and separation change in a well-determined way, which will be explained in Chapter~4.
The evolution of close binaries in this more realistic approach was first calculated, independently of one another, by \citet{pac66}, \citet{kw67}, and \citet{pla67}.
Their work was the foundation of all subsequent work on the evolution of close binary systems.  
Furthermore, the discovery of the first celestial X-ray source in 1962 and of the X-ray binaries in 1971--1972 --- which earned Riccardo Giacconi the 2002 Physics Nobel Prize --- has given a great stimulus to the research in this field. The X-ray binaries consist of a normal star together with a compact object: a NS or a BH, 
which are the end states of evolution of massive stars. The X-rays from these binaries, which are observed with space-borne detectors, are generated by the accretion of matter onto the compact star, which is captured from the outer layers of the companion star. During the accretion this gas, falling inwards in the extremely strong gravitational field of the compact star, is heated by the release of gravitational potential energy to temperatures above $10^6\;{\rm K}$, causing it to emit X-rays. 

Without the occurrence of extensive mass transfer from the original primary star (the progenitor of the compact object) to the secondary star, most of these systems could not have survived the SN explosion of the primary star in which the NS or BH was formed \citep{vhh72,ty73a}, because the probability of the post-SN orbit to remain bound depends on the relative mass loss from the system during the SN (see e.g. Section~4.3.10 and Chapter 13).

Since the early 1970s, the realization of the importance of accretion of matter onto a compact star (NS, BH or white dwarf [WD]) as an energy source in many types of binaries, ranging from X-ray binaries to cataclysmic variables (CVs) and symbiotic stars, has been a further important source of inspiration for new research on the structure and evolution of close binary systems.
The discoveries since 1974 of binary radio pulsars, of which at least 20 are double NSs \citep[DNSs; for a review, see][]{tkf+17}, have revealed many interesting properties, including the many relativistic effects that are measurable in them with unprecedented precision \citep[e.g.][]{tw89,tay92,kk16}. 

These discoveries have created a new and fundamental branch of relativistic binary star astrophysics, which among other things has produced the most accurate measurements of masses of any stellar objects so far (Chapter~14). The measured rate of orbital decay of the first-discovered DNS, PSR~B1913+16, is in exact agreement with the decay rate predicted from the emission of GWs according to the general theory of relativity. The highly precise detection of this and other relativistic effects earned the discoverers of this binary pulsar, Russell Hulse and Joseph Taylor, the 1993 Physics Nobel Prize. Later-discovered DNSs, particularly the double pulsar system PSR~J0737$-$3039, have further refined the up to five tests of relativity allowed by these systems (Section~14.7.1) to almost incredible precision \citep{wex14,ksm+21}. 

The amazing discoveries of the merger event of a double BH, starting with GW150914 in September~2015 \citep{aaa+16a}, and of a DNS, in August~2017 \citep{aaa+17c}, have revealed the ultimate final destiny of a massive binary star system and demonstrated the production of strong bursts of GWs, observable on Earth. This earned the LIGO pioneers Rainer Weiss, Kip Thorne and Barry Barish the 2017 Physics Nobel Prize. 
But how do ordinary stars born in a binary system end up as two NSs or BHs, that finish as a final single BH remnant?
To answer this question, we need to follow the binary system through a long chain of exotic binary interactions (Fig.~1.1), involving mass transfer between the stellar components, SNe, and relativistic effects. 
That the orbits of evolved massive stars in binaries will shrink to the very small sizes observed for the double compact objects, was predicted before these objects were discovered \citep{vhdl73}. 

\begin{figure*}[!t]
\vspace{0.5cm}
  \begin{center}
  \includegraphics[width=1.00\textwidth, angle=0]{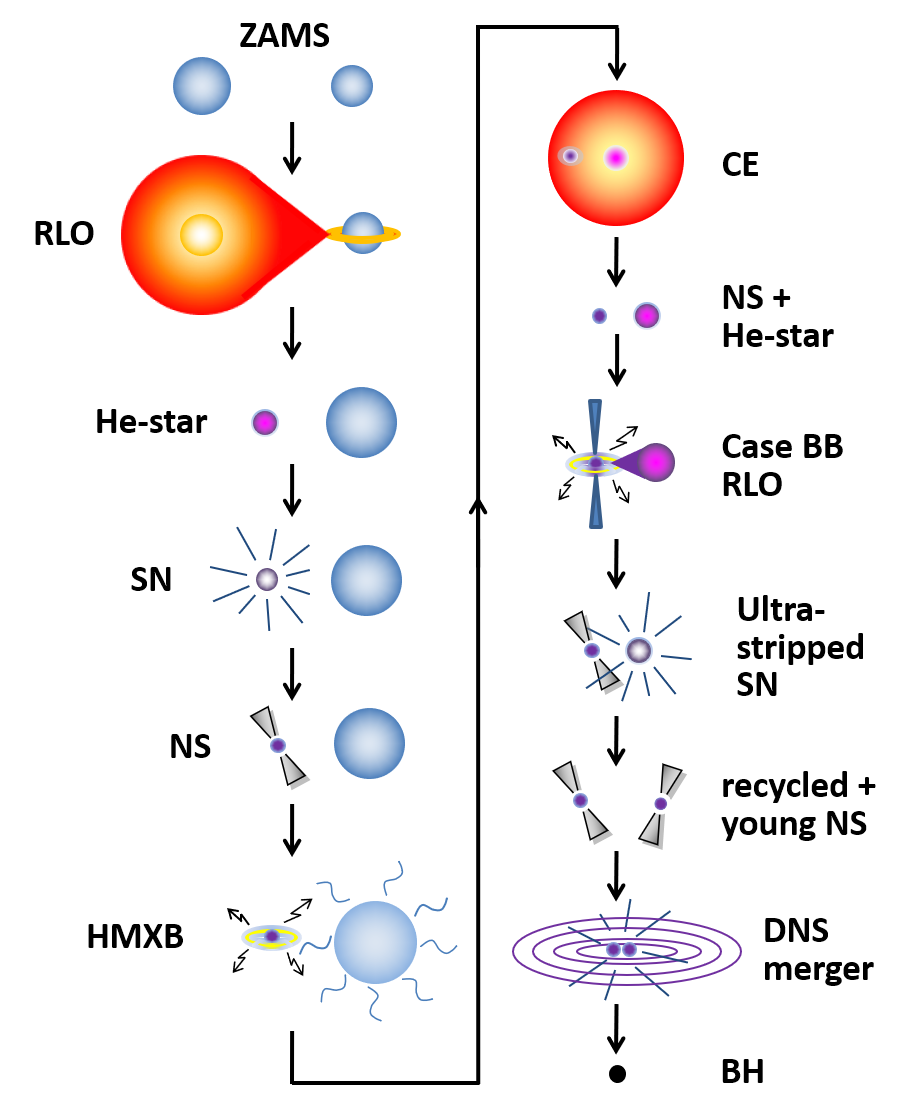}
\caption{Formation model of a close double neutron star (DNS) system as final product of the evolution of a massive close binary. The DNS system may eventually merge and leave a solitary BH remnant. The same model, scaled up to higher initial stellar masses, is one of the main scenarios to explain the formation of double BHs (see Chapters~10, 12, and 15). \citep[After][]{tkf+17}.}
  \label{fig:vdh_cartoon_DNS}
  \end{center}
\end{figure*}

\begin{figure*}[!t]
\vspace{-0.5cm}
  \begin{center}
  \includegraphics[width=1.15\textwidth, angle=0]
{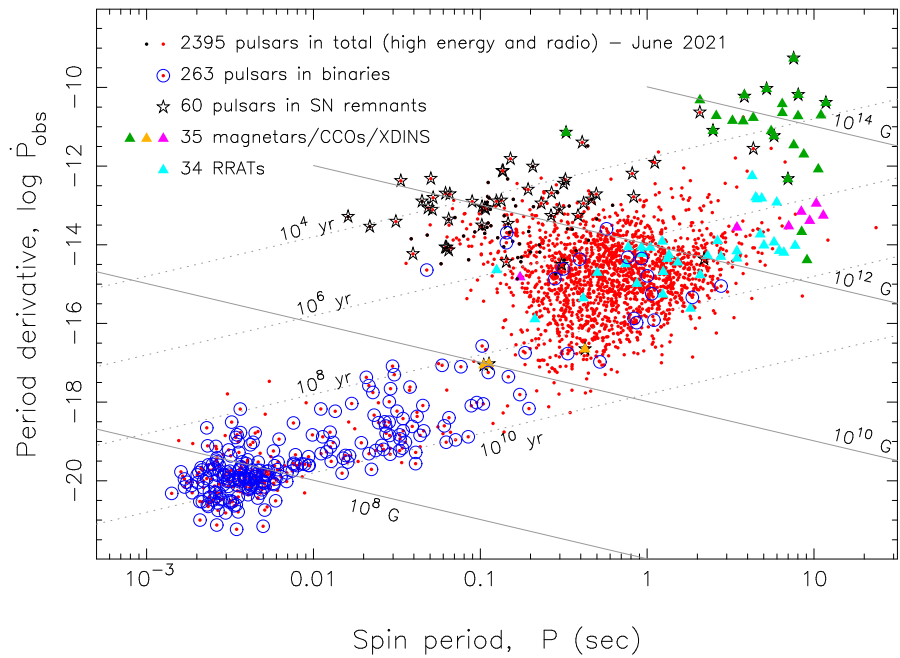}
\caption{Observed population of $\sim\!2400$ radio and high-energy pulsars with measured values of both spin period ($P$) and period derivative ($\dot{P}$). The variety of pulsars has flourished immensely since their discovery in 1967 and continues to be a major science driver in modern astrophysics. Binary pulsars (blue circles) dominate among the fast-spinning millisecond pulsars. CCOs are central compact objects of SN remnants, XDINSs are X-ray dim isolated NSs, and RRATs are rotating radio transients (see Chapter 14).  Data taken from the {\em ATNF Pulsar Catalogue} in June~2021
    \citep[][\url{https://www.atnf.csiro.au/research/pulsar/psrcat}]{mhth05}.}
  \label{fig:PPdot}
  \end{center}
\end{figure*}

Radio pulsars are some of the most intriguing astrophysical objects, among which a sizable fraction of binaries is found with quite special characteristics that give important information on binary evolution. 
To understand the continuously growing diversity of observed radio pulsars (see Fig.~1.2), it is necessary to link their properties to the stellar and binary evolution of their progenitors \citep{bv91}. 

The detection of close binary pulsars, and of the merger events of double BHs and NSs have further increased the interest in the evolution of binary systems. These discoveries demonstrated that, even though binaries may have undergone several stages of mass transfer during their evolution, where up to 90\% of their original mass and $>95$\% of their original orbital angular momentum is lost from the system, and despite having experienced two SN explosions, the two stars might still survive as a (very close) binary system with two compact objects. 

The existence of the close DNSs and close double BHs has also demonstrated that a precise knowledge of the physics of binary evolution is of vital importance for understanding fundamental astrophysics as diverse as the generation of the strongest bursts of GW radiation, the production of gamma-ray bursts, and the synthesis of heavy r-process (or rapid neutron capture) elements in the Universe. The latter was predicted to be produced by the merging of DNSs or NS+BH binaries \citep{ls76} and expected to be observable as a so-called kilonova optical-infrared eruption \citep[]{mmd+10,ber14}. 
          
These predictions have been beautifully confirmed by the spectroscopic study of the optical-infrared transient that accompanied the DNS merger event \linebreak GW170817 \citep[]{aaa+17c,aaa+17d,aaa+17e}. 
The mergers of double compact object binaries with at least one NS also had been predicted to produce gamma-ray bursts \citep[]{pac86,elps89}, which was confirmed by the Fermi and INTEGRAL missions by the detection of a short gamma-ray burst, following within two seconds after GW170817 \cite[]{aaa+17f,sfk+17}. The occurrence of an electromagnetic counterpart --- the kilonova --- is a great advantage of a DNS merger or a NS+BH merger, relative to the mergers of double BHs, because this allows one to determine the place of origin of the merger on the sky with sub-arcsecond precision. In the case of GW170817, this place turned out to be in the lenticular (S0) galaxy, NGC~4993 at a distance of 40~Mpc \citep[]{cfk+17,sha+17}. The discovery, in this way, of the DNS merger GW170817 at once provided the solution to the nature of a variety of key astrophysical phenomena.    

In the past decades it was realized that the SNe of Types~Ia, Ib and Ic \linebreak (characterized by the absence of hydrogen in their spectra) most likely are related to the evolution of binary systems (see Section~2.10). That these three types together form about half of all SNe shows how important knowledge of binary star evolution is for a general understanding of observational stellar evolution processes.
         
Finally, in the past decades it was discovered that binary evolution has affected a large variety of low- and intermediate-mass stars with peculiar abundances of chemical elements. Examples are the barium stars, which are G and K giants with masses up to a few solar masses and an overabundance of barium and other s-process elements (products of slow neutron-capture processes \citep[]{bk51,bj88}. They are wide binaries with eccentric orbits and orbital periods ranging from approximately 100 to 10\,000~days \citep{bj88}. The binary nature of these stars, of blue stragglers in star clusters, of extremely metal-depleted post AGB stars, of carbon enhanced metal poor (CEMP) stars, and of over 80\% of nuclei of planetary nebulae was discovered in the past decades. These discoveries show that the evolution of $\sim\!50$\% of all low-and intermediate-mass stars is expected to be affected by binary evolution \citep{api15}. Because lower-mass stars are far more abundant in galaxies than massive stars are, the products of low- and intermediate-mass binaries are expected to vastly outnumber the products of the evolution of massive binaries. However, in this book, we will concentrate mainly on the physics and evolution of the more massive binaries that produce NSs and BHs and on the origin of Type~Ia SNe, which are caused by exploding WDs and are of crucial importance for cosmology. 
 
In the last decades it was also realized that stars are found often even in triple systems or higher-order multiple systems.  
Observational estimates suggest that approximately $20-30$\% of all binary stars are members of triple systems \citep{ttsu06,rdl+13}. 
These systems may remain bound with a long-term stability in a hierarchical structure (a close inner binary with a third star in relatively distant orbit). Since the 1970s, a number of stability criteria for a triple system have been proposed \citep[see][for an overview]{mik08} that enable predictions for their long-term stability. 
\citet{it99} estimated that in 
$\sim\!70$\% of the triple systems, the inner binary is close enough that the most massive star will evolve to fill its Roche~lobe. 
Furthermore, in $\sim\!15$\% of the triple star systems, the outer third (tertiary) star may even also fill its
Roche~lobe at some point, possibly leading to disintegration or production of rare configurations with three degenerate objects in the same system. In 2014, \citet{rsa+14} reported the remarkable discovery of a triple-system pulsar (PSR~J0337+1715), which is exactly the first example of such an exotic system --- a NS orbited by two WDs. \citep[For a formation and evolution scenario for this complex system, which must have survived a SN explosion and at least three stages of mass transfer between the stellar components, see our model in][]{tv14}.
Besides specific triple-star interactions, such as Kozai-Lidov resonances \citep{koz62,lid62} and the above-mentioned dynamical stability considerations, most interactions between stars in triples are similar to the interactions between binary stars. For this reason, we focus only on binary stars in this book\footnote{Formation of stellar binaries via triple- or quadruple-star dynamical interactions in dense cluster, however, are discussed in Chapter~12. For a recent review on the evolution of destabilized triple systems, see \citet{tbp22}.}.

The 2020s and 2030s are expected to reveal a large number of discoveries of new double compact object systems, as well as their progenitors and merger remnants. This field of astrophysics is strongly driven by investments in new big-science instruments. The Square-Kilometer Array (SKA) is expected to increase the number of known radio pulsars by a factor of 5 to 10, thus resulting in a total of $>\!100$ known DNS systems \citep{kbk+15}. The Five-hundred-meter Aperture Spherical Telescope (FAST) is also expected to contribute a significant number of new radio pulsars \citep{slk+09,nlp+11}, including new discoveries of pulsar binaries. High-mass X-ray binaries (HMXBs), the anticipated progenitors of double compact object systems containing NSs and BHs, are continuously being discovered with ongoing X-ray missions \citep[INTEGRAL, Swift, XMM-Newton, and Chandra, see e.g.][]{cha13}. New and upcoming space-borne X-ray telescopes such as eXTP, STROBE-X, and Athena are expected to produce further discoveries of these systems. Hence, we are currently in an epoch when a large wealth of new information on exotic binaries is becoming available. In light of this, it is important to explore and understand the formation and evolution of such binary systems in more detail.
Earlier textbooks on binary evolution are those of \citet{slv94}, and \citet{egg06}, to which we refer for further reading. For an earlier book on physics and evolution of relativistic objects in binaries, we refer to \citet{ccg+09}.

This book is organized as follows.  
In Chapter~2, we give a brief history of the discovery of the different types of binary systems and, where appropriate, summarize their importance in modern astrophysics. 
In Chapter~3, we consider how the orbital parameters of spectroscopic and eclipsing binaries are measured and how from these measurements, information is obtained about the masses and radii of the stars. We give an overview of the thus derived masses of stars of different spectral types. 

In Chapter~4, we consider basic aspects of the celestial mechanics of binary systems, the meaning and limitations of the Roche-lobe concept, as well as the changes in orbital period and binary separation that are induced by various processes of mass loss and mass transfer in binary systems.
We first consider the somewhat idealized ``conservative'' evolution, in which the total mass and orbital angular momentum of the binary are expected to be conserved during the evolution, followed by the more realistic and exotic types of close binary evolution, so-called non-conservative evolution, in which large losses of mass and orbital angular momentum from the systems are taken into account. 
A treatment of common envelopes and the orbital evolution during the dynamically unstable in-spiral phase are described as well.
Finally, we briefly discuss the Eddington accretion limit as well as accretion disks.

In Chapter~5, we describe the observed properties and the general classifications of the various types of interacting binary systems, that do not contain NSs or BHs, concentrating mostly on systems in which at least one component is an evolved star, that is, a (sub)giant, or a WD.

In Chapters~6 and 7, we describe the observed properties of X-ray binaries: high-mass X-ray binaries (HMXBs) as well as low-mass X-ray binaries (LMXBs), including mass determination of the accreting NSs and BHs.
Regarding HMXBs, we discuss Be-star X-ray binaries, supergiant X-ray binaries, and, for example, stellar wind accretion and the Corbet diagram. We also discuss the recently discovered class of pulsating ultra-luminous X-ray sources. For LMXBs, we discuss the various types, including the systems with BHs and the symbiotic X-ray binaries with accreting NSs.

In Chapter~8, we give an overview of the evolution of single stars (with a special focus on the final evolution of massive stars).
In Chapter~9, we apply this knowledge to the evolution of binaries in general. Here, we also discuss in detail the various cases of mass transfer (including mass loss) and orbital stability analysis, and we end with a comparison between the outcomes of single versus binary star evolution.

This knowledge is crucial for understanding the formation and evolution of X-ray binaries, which are the subjects of Chapters~10 (HMXBs) and 11 (LMXBs). We discuss the final stages of HMXBs (including Wolf-Rayet star binaries), with or without a common envelope, leading to the formation of double NS/BH binaries. 
For LMXBs and CVs, we also discuss the mechanisms driving the mass transfer in LMXBs and CVs. These concern the internal evolution of the companion star, as well as the loss of orbital angular momentum due to emission of GWs and/or a magnetically coupled stellar wind, or a combination of these. We also discuss the final mass-transfer stage from WDs in very tight binaries, in the so-called AM~Canum Venaticorum (double WD) systems and the ultra-compact X-ray binary sources (typically WD+NS systems).

Binaries with compact objects can also be formed by the dynamical evolution of dense star clusters. This is the subject of Chapter~12.

Chapter~13 concerns SNe in binaries. We first discuss the evolution leading to the thermonuclear Type~Ia SNe, triggered by the accretion of matter by a WD, and we also consider accretion-induced collapse of massive WDs. We subsequently discuss the origin of Type~Ib/Ic SNe, and the evidence derived from theoretical computations of the late stages of close binary evolution, including ultra-stripped SNe, combined with observations of SN light curves and known Galactic post-SN DNS systems. We discuss the evidence for momentum kicks imparted onto compact objects during various SN events (iron core collapse and electron-capture SNe) and examine the kinematical effects of these kicks on the resulting compact-object binaries. 

In Chapter~14, using the results of the previous chapters, the final stages of the formation of binary and millisecond radio pulsars is studied; in other words, the transition from X-ray binaries to radio pulsars. We review the recycling of pulsars in detail, including the accretion torques at work. We review the rich diversity of resulting millisecond pulsar binaries and their component masses (theoretical expectations and measurements via relativistic effects). Finally, we revisit the formation of DNS systems.

Chapter~15 is devoted to GW astrophysics. We cover the basic physics of GW emission and detection. We discuss formation channels of both high-frequency (LIGO--Virgo--KAGRA--IndIGO) and low-frequency (LISA and TianQin) GW sources. We review the signals expected from extragalactic merging double BHs and DNSs, including the electromagnetic counterparts of these mergers, and the Galactic WD and NS binaries as continuous GW emission sources. The different models for the formation of the double BHs are discussed in depth in light of the latest observations of ``ordinary'' and exotic events from the LIGO--Virgo--KAGRA network.

Chapter~16 discusses the subject of binary population synthesis with an emphasis on methodology and statistics. Two examples are highlighted that illustrate a synthetic open star cluster population of binaries and the differences between estimates of empirical versus theoretical DNS merger rates. 

 


\mbox{}
\section*{}
\addcontentsline{toc}{section}{REFERENCES}
\bibliographystyle{mn2e} 
\bibliography{book_arxiv.bib}

\begin{thebibliography}{}

\bibitem[\protect\citeauthoryear{{Abate}, {Pols}, {Izzard} \&
  {Karakas}}{{Abate} et~al.}{2015}]{api15}
{Abate} C.,  {Pols} O.~R.,  {Izzard} R.~G.,    {Karakas} A.~I.,  2015, \aap,
  581, A22

\bibitem[\protect\citeauthoryear{{Abbott}, {Abbott}, {Abbott}, {Abernathy},
  {Acernese}, {Ackley}, {Adams}, {Adams}, {Addesso}, {Adhikari} \& et
  al.}{{Abbott} et~al.}{2016}]{aaa+16a}
{Abbott} B.~P.,  {Abbott} R.,  {Abbott} T.~D.,  {Abernathy} M.~R.,  {Acernese}
  F.,  {Ackley} K.,  {Adams} C.,  {Adams} T.,  {Addesso} P.,  {Adhikari} R.~X.,
     et al. 2016, Physical Review Letters, 116, 061102

\bibitem[\protect\citeauthoryear{{Abbott}, {Abbott}, {Abbott}, {Acernese},
  {Ackley}, {Adams} \& {et~al.}}{{Abbott} et~al.}{2017a}]{aaa+17f}
{Abbott} B.~P.,  {Abbott} R.,  {Abbott} T.~D.,  {Acernese} F.,  {Ackley} K.,
  {Adams} C.,    {et~al.} 2017a, \apjl, 848, L13

\bibitem[\protect\citeauthoryear{{Abbott}, {Abbott}, {Abbott}, {Acernese},
  {Ackley}, {Adams} \& {et~al.}}{{Abbott} et~al.}{2017b}]{aaa+17c}
{Abbott} B.~P.,  {Abbott} R.,  {Abbott} T.~D.,  {Acernese} F.,  {Ackley} K.,
  {Adams} C.,    {et~al.} 2017b, Physical Review Letters, 119, 161101

\bibitem[\protect\citeauthoryear{{Abbott}, {Abbott}, {Abbott}, {Acernese},
  {Ackley}, {Adams} \& {et~al.}}{{Abbott} et~al.}{2017c}]{aaa+17e}
{Abbott} B.~P.,  {Abbott} R.,  {Abbott} T.~D.,  {Acernese} F.,  {Ackley} K.,
  {Adams} C.,    {et~al.} 2017c, \apjl, 848, L12

\bibitem[\protect\citeauthoryear{{Abbott}, {Abbott}, {Abbott}, {Acernese},
  {Ackley}, {Adams} \& {et~al.}}{{Abbott} et~al.}{2017d}]{aaa+17d}
{Abbott} B.~P.,  {Abbott} R.,  {Abbott} T.~D.,  {Acernese} F.,  {Ackley} K.,
  {Adams} C.,    {et~al.} 2017d, \apjl, 850, L40

\bibitem[\protect\citeauthoryear{{Abt} \& {Levy}}{{Abt} \&
  {Levy}}{1976}]{abtl76}
{Abt} H.~A.,  {Levy} S.~G.,  1976, \apjs, 30, 273

\bibitem[\protect\citeauthoryear{{Berger}}{{Berger}}{2014}]{ber14}
{Berger} E.,  2014, \araa, 52, 43

\bibitem[\protect\citeauthoryear{Bhattacharya \& {van den Heuvel}}{Bhattacharya
  \& {van den Heuvel}}{1991}]{bv91}
Bhattacharya D.,  {van den Heuvel} E. P.~J.,  1991, Physics Reports, 203, 1

\bibitem[\protect\citeauthoryear{{Bidelman} \& {Keenan}}{{Bidelman} \&
  {Keenan}}{1951}]{bk51}
{Bidelman} W.~P.,  {Keenan} P.~C.,  1951, \apj, 114, 473

\bibitem[\protect\citeauthoryear{{Boffin} \& {Jorissen}}{{Boffin} \&
  {Jorissen}}{1988}]{bj88}
{Boffin} H.~M.~J.,  {Jorissen} A.,  1988, \aap, 205, 155

\bibitem[\protect\citeauthoryear{{Bonnell}, {Bate} \& {Vine}}{{Bonnell}
  et~al.}{2003}]{bbv03}
{Bonnell} I.~A.,  {Bate} M.~R.,    {Vine} S.~G.,  2003, \mnras, 343, 413

\bibitem[\protect\citeauthoryear{{Chaty}}{{Chaty}}{2013}]{cha13}
{Chaty} S.,  2013, Advances in Space Research, 52, 2132

\bibitem[\protect\citeauthoryear{{Chini}, {Nasseri}, {Hoffmeister}, {Buda} \&
  {Barr}}{{Chini} et~al.}{2011}]{cnh+11}
{Chini} R.,  {Nasseri} A.,  {Hoffmeister} V.~H.,  {Buda} L.~S.,    {Barr} A.,
  2011, in {Schmidtobreick} L.,  {Schreiber} M.~R.,   {Tappert} C.,  eds,
  Evolution of Compact Binaries Vol.~447 of Astronomical Society of the Pacific
  Conference Series, {Most High-Mass Stars are Born as Twins}.
p.~67

\bibitem[\protect\citeauthoryear{{Colpi}, {Casella}, {Gorini}, {Moschella} \&
  {Possenti}}{{Colpi} et~al.}{2009}]{ccg+09}
{Colpi} M.,  {Casella} P.,  {Gorini} V.,  {Moschella} U.,    {Possenti} A.,
  2009, {Physics of Relativistic Objects in Compact Binaries: From Birth to
  Coalescence}.
Vol.~359

\bibitem[\protect\citeauthoryear{{Coulter}, {Foley}, {Kilpatrick} \&
  {et~al.}}{{Coulter} et~al.}{2017}]{cfk+17}
{Coulter} D.~A.,  {Foley} R.~J.,  {Kilpatrick} C.~D.,    {et~al.} 2017,
  Science, 358, 1556

\bibitem[\protect\citeauthoryear{{Crawford}}{{Crawford}}{1955}]{cra55}
{Crawford} J.~A.,  1955, \apj, 121, 71

\bibitem[\protect\citeauthoryear{{Eggleton}}{{Eggleton}}{2006}]{egg06}
{Eggleton} P.,  2006, {Evolutionary Processes in Binary and Multiple Stars}

\bibitem[\protect\citeauthoryear{{Eichler}, {Livio}, {Piran} \&
  {Schramm}}{{Eichler} et~al.}{1989}]{elps89}
{Eichler} D.,  {Livio} M.,  {Piran} T.,    {Schramm} D.~N.,  1989, \nat, 340,
  126

\bibitem[\protect\citeauthoryear{{Iben} Jr. \& {Tutukov}}{{Iben} \&
  {Tutukov}}{1999}]{it99}
{Iben} Jr. I.,  {Tutukov} A.~V.,  1999, \apj, 511, 324

\bibitem[\protect\citeauthoryear{{Kaspi} \& {Kramer}}{{Kaspi} \&
  {Kramer}}{2016}]{kk16}
{Kaspi} V.~M.,  {Kramer} M.,  2016, in {Blandford} R.,  {Gross} D.,   {Sevrin}
  A.,  eds, Proceedings of the 26th Solvay Conference on Physics Vol.~26 of
  Astrophysics and Cosmology, World Scientiﬁc Publishing Comp., {Radio
  Pulsars: The Neutron Star Population \& Fundamental Physics}.
pp 21--70

\bibitem[\protect\citeauthoryear{{Keane}, {Bhattacharyya}, {Kramer} \&
  {et~al.}}{{Keane} et~al.}{2015}]{kbk+15}
{Keane} E.,  {Bhattacharyya} B.,  {Kramer} M.,    {et~al.} 2015, Advancing
  Astrophysics with the Square Kilometre Array (AASKA14), p.~40

\bibitem[\protect\citeauthoryear{{Kippenhahn} \& {Weigert}}{{Kippenhahn} \&
  {Weigert}}{1967}]{kw67}
{Kippenhahn} R.,  {Weigert} A.,  1967, \zap, 65, 251

\bibitem[\protect\citeauthoryear{{Kozai}}{{Kozai}}{1962}]{koz62}
{Kozai} Y.,  1962, \aj, 67, 591

\bibitem[\protect\citeauthoryear{{Kramer}, {Stairs}, {Manchester}, {Wex},
  {Deller}, {Coles} \& {et~al.}}{{Kramer} et~al.}{2021}]{ksm+21}
{Kramer} M.,  {Stairs} I.~H.,  {Manchester} R.~N.,  {Wex} N.,  {Deller} A.~T.,
  {Coles} W.~A.,    {et~al.} 2021, Physical Review X, 11, 041050

\bibitem[\protect\citeauthoryear{{Lattimer} \& {Schramm}}{{Lattimer} \&
  {Schramm}}{1976}]{ls76}
{Lattimer} J.~M.,  {Schramm} D.~N.,  1976, \apj, 210, 549

\bibitem[\protect\citeauthoryear{{Lidov}}{{Lidov}}{1962}]{lid62}
{Lidov} M.~L.,  1962, \planss, 9, 719

\bibitem[\protect\citeauthoryear{{Manchester}, {Hobbs}, {Teoh} \&
  {Hobbs}}{{Manchester} et~al.}{2005}]{mhth05}
{Manchester} R.~N.,  {Hobbs} G.~B.,  {Teoh} A.,    {Hobbs} M.,  2005, \aj, 129,
  1993

\bibitem[\protect\citeauthoryear{{Metzger}, {Mart{\'{\i}}nez-Pinedo}, {Darbha},
  {Quataert}, {Arcones}, {Kasen}, {Thomas}, {Nugent}, {Panov} \&
  {Zinner}}{{Metzger} et~al.}{2010}]{mmd+10}
{Metzger} B.~D.,  {Mart{\'{\i}}nez-Pinedo} G.,  {Darbha} S.,  {Quataert} E.,
  {Arcones} A.,  {Kasen} D.,  {Thomas} R.,  {Nugent} P.,  {Panov} I.~V.,
  {Zinner} N.~T.,  2010, \mnras, 406, 2650

\bibitem[\protect\citeauthoryear{{Mikkola}}{{Mikkola}}{2008}]{mik08}
{Mikkola} S.,  2008, in {Hubrig} S.,  {Petr-Gotzens} M.,   {Tokovinin} A.,
  eds, Multiple Stars Across the H-R Diagram {Dynamics and Stability of Triple
  Stars}.
p.~11

\bibitem[\protect\citeauthoryear{{Morton}}{{Morton}}{1960}]{mor60}
{Morton} D.~C.,  1960, \apj, 132, 146

\bibitem[\protect\citeauthoryear{{Nan}, {Li}, {Jin}, {Wang}, {Zhu}, {Zhu},
  {Zhang}, {Yue} \& {Qian}}{{Nan} et~al.}{2011}]{nlp+11}
{Nan} R.,  {Li} D.,  {Jin} C.,  {Wang} Q.,  {Zhu} L.,  {Zhu} W.,  {Zhang} H.,
  {Yue} Y.,    {Qian} L.,  2011, International Journal of Modern Physics D, 20,
  989

\bibitem[\protect\citeauthoryear{{Paczy{\'n}ski}}{{Paczy{\'n}ski}}{1966}]{pac66}
{Paczy{\'n}ski} B.,  1966, \actaa, 16, 231

\bibitem[\protect\citeauthoryear{{Paczynski}}{{Paczynski}}{1986}]{pac86}
{Paczynski} B.,  1986, \apjl, 308, L43

\bibitem[\protect\citeauthoryear{{Plavec}}{{Plavec}}{1967}]{pla67}
{Plavec} M.,  1967, Bulletin of the Astronomical Institutes of Czechoslovakia,
  18, 253

\bibitem[\protect\citeauthoryear{{Ransom}, {Stairs}, {Archibald}, {Hessels},
  {Kaplan}, {van Kerkwijk} \& {et~al.}}{{Ransom} et~al.}{2014}]{rsa+14}
{Ransom} S.~M.,  {Stairs} I.~H.,  {Archibald} A.~M.,  {Hessels} J.,  {Kaplan}
  D.~L.,  {van Kerkwijk} M.~H.,    {et~al.} 2014, \nat, 505, 520

\bibitem[\protect\citeauthoryear{{Rappaport}, {Deck}, {Levine}, {Borkovits},
  {Carter}, {El Mellah}, {Sanchis-Ojeda} \& {Kalomeni}}{{Rappaport}
  et~al.}{2013}]{rdl+13}
{Rappaport} S.,  {Deck} K.,  {Levine} A.,  {Borkovits} T.,  {Carter} J.,  {El
  Mellah} I.,  {Sanchis-Ojeda} R.,    {Kalomeni} B.,  2013, \apj, 768, 33

\bibitem[\protect\citeauthoryear{{Sana}, {de Mink}, {de Koter}, {Langer},
  {Evans}, {Gieles}, {Gosset}, {Izzard}, {Le Bouquin} \& {Schneider}}{{Sana}
  et~al.}{2012}]{sdd+12}
{Sana} H.,  {de Mink} S.~E.,  {de Koter} A.,  {Langer} N.,  {Evans} C.~J.,
  {Gieles} M.,  {Gosset} E.,  {Izzard} R.~G.,  {Le Bouquin} J.-B.,
  {Schneider} F.~R.~N.,  2012, Science, 337, 444

\bibitem[\protect\citeauthoryear{{Savchenko}, {Ferrigno}, {Kuulkers} \&
  {et~al.}}{{Savchenko} et~al.}{2017}]{sfk+17}
{Savchenko} V.,  {Ferrigno} C.,  {Kuulkers} E.,    {et~al.} 2017, \apjl, 848,
  L15

\bibitem[\protect\citeauthoryear{{Shore}, {Livio} \& {van den Heuvel}}{{Shore}
  et~al.}{1994}]{slv94}
{Shore} S.~N.,  {Livio} M.,    {van den Heuvel} E.~P.~J.,  1994, {Interacting
  binaries}

\bibitem[\protect\citeauthoryear{{Smits}, {Lorimer}, {Kramer}, {Manchester},
  {Stappers}, {Jin}, {Nan} \& {Li}}{{Smits} et~al.}{2009}]{slk+09}
{Smits} R.,  {Lorimer} D.~R.,  {Kramer} M.,  {Manchester} R.,  {Stappers} B.,
  {Jin} C.~J.,  {Nan} R.~D.,    {Li} D.,  2009, \aap, 505, 919

\bibitem[\protect\citeauthoryear{{Soares-Santos}, {Holz}, {Annis} \&
  {et~al.}}{{Soares-Santos} et~al.}{2017}]{sha+17}
{Soares-Santos} M.,  {Holz} D.~E.,  {Annis} J.,    {et~al.} 2017, \apjl, 848,
  L16

\bibitem[\protect\citeauthoryear{{Tauris}, {Kramer}, {Freire}, {Wex}, {Janka},
  {Langer}, {Podsiadlowski}, {Bozzo}, {Chaty}, {Kruckow}, {van den Heuvel},
  {Antoniadis}, {Breton} \& {Champion}}{{Tauris} et~al.}{2017}]{tkf+17}
{Tauris} T.~M.,  {Kramer} M.,  {Freire} P.~C.~C.,  {Wex} N.,  {Janka} H.-T.,
  {Langer} N.,  {Podsiadlowski} P.,  {Bozzo} E.,  {Chaty} S.,  {Kruckow} M.~U.,
   {van den Heuvel} E.~P.~J.,  {Antoniadis} J.,  {Breton} R.~P.,    {Champion}
  D.~J.,  2017, \apj, 846, 170

\bibitem[\protect\citeauthoryear{{Tauris} \& {van den Heuvel}}{{Tauris} \& {van
  den Heuvel}}{2014}]{tv14}
{Tauris} T.~M.,  {van den Heuvel} E.~P.~J.,  2014, \apjl, 781, L13

\bibitem[\protect\citeauthoryear{{Taylor}}{{Taylor}}{1992}]{tay92}
{Taylor} J.~H.,  1992, Royal Society of London Philosophical Transactions
  Series A, 341, 117

\bibitem[\protect\citeauthoryear{{Taylor} \& {Weisberg}}{{Taylor} \&
  {Weisberg}}{1989}]{tw89}
{Taylor} J.~H.,  {Weisberg} J.~M.,  1989, \apj, 345, 434

\bibitem[\protect\citeauthoryear{{Tokovinin}, {Thomas}, {Sterzik} \&
  {Udry}}{{Tokovinin} et~al.}{2006}]{ttsu06}
{Tokovinin} A.,  {Thomas} S.,  {Sterzik} M.,    {Udry} S.,  2006, \aap, 450,
  681

\bibitem[\protect\citeauthoryear{{Toonen}, {Boekholt} \& {Portegies
  Zwart}}{{Toonen} et~al.}{2022}]{tbp22}
{Toonen} S.,  {Boekholt} T.~C.~N.,    {Portegies Zwart} S.,  2022, \aap, 661,
  A61

\bibitem[\protect\citeauthoryear{{Tutukov} \& {Yungelson}}{{Tutukov} \&
  {Yungelson}}{1973}]{ty73a}
{Tutukov} A.,  {Yungelson} L.,  1973, Nauchnye Informatsii, 27, 70

\bibitem[\protect\citeauthoryear{{van den Heuvel}}{{van den
  Heuvel}}{1994}]{vdh94a}
{van den Heuvel} E.~P.~J.,  1994, in {Shore} S.~N.,  {Livio} M.,  {van den
  Heuvel} E.~P.~J.,  {Nussbaumer} H.,   {Orr} A.,  eds, Saas-Fee Advanced
  Course 22: Interacting Binaries {Interacting binaries: topics in close binary
  evolution.}.
pp 263--474

\bibitem[\protect\citeauthoryear{{van den Heuvel} \& {De Loore}}{{van den
  Heuvel} \& {De Loore}}{1973}]{vhdl73}
{van den Heuvel} E.~P.~J.,  {De Loore} C.,  1973, \aap, 25, 387

\bibitem[\protect\citeauthoryear{{van den Heuvel} \& {Heise}}{{van den Heuvel}
  \& {Heise}}{1972}]{vhh72}
{van den Heuvel} E.~P.~J.,  {Heise} J.,  1972, Nature Physical Science, 239, 67

\bibitem[\protect\citeauthoryear{{Wex}}{{Wex}}{2014}]{wex14}
{Wex} N.,  2014, in {Kopeikein} S.~M.,  ed., Brumberg Festschrift Published by
  de Gruyter, Berlin, 2014, {Testing Relativistic Gravity with Radio Pulsars}.
pp 1--80

\end{thebibliography}

\end{document}